\input harvmac

\Title{\vbox{\baselineskip12pt\hbox{}}}
{\vbox{\centerline{Reply to: Comment on ``Quantum Optimization}
\vskip2pt\centerline{for Combinatorial Searches''}}}
\centerline{C. A. Trugenberger}
\centerline{InfoCodex SA, av. Louis-Casa\"i 18, CH-1209 Geneva, Switzerland}
\centerline{Theory Division, CERN,CH-1211 Geneva 23, Switzerland}
\centerline{ca.trugenberger@InfoCodex.com}

\vskip .3in
\centerline{Abstract}

This is my reply to Zalka and Brun's criticism of my recent paper
on quantum optimization heuristics. Essentially, this criticism is shown
to be utterly irrelevant.

\vskip .3in

In a recent publication \ref\cat{C.A. Trugenberger, {\it Quantum Optimization
for Combinatorial Searches}, New Journal of Physics {\bf 4} (2002) 26.1-26.7.}\
I proposed a quantum optimization heuristic for combinatorial search problems. The
basic idea is to generate a quantum superposition  of search states such that the
corresponding probability distribution  for measurement outcomes is strongly
peaked on low-cost states, in analogy to a thermal Boltzmann distribution, and to
apply then standard thermodynamic reasoning, as in simulated annealing.

This procedure has been criticized by Zalka and Brun \ref\zal{C. Zalka and T.
Brun, {\it Comment on ``Quantum Optimization for Combinatorial Searches''},
quant-ph/0206081.}\ on the ground that the probability of measurement of any fixed
states $I^k$ is small (last but one paragraph of their Comment \zal).

Unfortunately, these authors have forgotten that, once the desired {\it thermal}
quantum superposition has been generated (which is guarnteed by obtaining a
particular outcome from the measurement of an auxiliary register), it is not at
all the absolute probabilities that matter but only the relative ones. After all,
nobody would discard statistical mechanics on the ground that the canonical
partition function, providing the normalization of occupation probabilities, is
large for a large number of degrees of freedom. In other words, even if the
absolute probability of measuring one particular state $I^k$ with cost $C(I^k)$ is
small, what matters is that the probability of measuring any other state $I^j$ with
$C(I^j) > C(I^k)$ is even much smaller, actually infinitely smaller when the
effective temperature $t$ approaches zero. As I point out in my paper, the really
important question is another one, and namely how low one has to choose the
temperature in order to achive  a given accuracy. Since the effective temperature
is set by the inverse of the number $b$ of auxiliary qbits, increasing the
accuracy requires more computational load. But this tradeoff between accuracy
and computational load is common to all optimization heuristics, which are, after
all, approximation techniques.

So, while Zalka and Brun's comment contains mathematically correct formulas, it is
also totally irrelevant to the problem at hand.

\listrefs
\end